# Enhancement of perpendicular magnetic anisotropy and Dzyaloshinskii-Moriya interaction in thin ferromagnetic films by atomic-scale modulation of interfaces

A. S. Samardak,[1,5*] A. V. Davydenko,[1] A. G. Kolesnikov,[1] A. Yu. Samardak,[1] A. G. Kozlov[1], Bappaditya Pal,[1] A. V. Ognev,[1] A. V. Sadovnikov,[2,3] S. A. Nikitov,[2,3] A. V. Gerasimenko,[4] In Ho Cha,[6] Yong Jin Kim,[6] Gyu Won Kim,[6] Oleg A. Tretiakov,[7,8] Young Keun Kim[6*]

[1]*School of Natural Sciences, Far Eastern Federal University, Vladivostok 690950, Russia*

[2]*Laboratory "Metamaterials," Saratov State University, Saratov 410012, Russia*

[3]*Kotel'nikov Institute of Radioengineering and Electronics, Russian Academy of Sciences, Moscow 125009, Russia*

[4]*Institute of Chemistry, Far East Branch, Russian Academy of Sciences, Vladivostok 690022, Russia*

[5]*National Research South Ural State University, Chelyabinsk 454080, Russia*

[6]*Department of Materials Science and Engineering, Korea University, Seoul 02841, Republic of Korea*

[7] *School of Physics, The University of New South Wales, Sydney 2052, Australia*

[8] *National University of Science and Technology ``MISiS'', Moscow 119049, Russia*

*e-mails: samardak.as@dvfu.ru; ykim97@korea.ac.kr

**Abstract**

To stabilize the non-trivial spin textures, e.g., skyrmions or chiral domain walls in ultrathin magnetic films, an additional degree of freedom such as the interfacial Dzyaloshinskii–Moriya interaction (IDMI) must be induced by the strong spin-orbit coupling (SOC) of a stacked heavy metal layer. However, advanced approaches to simultaneously control IDMI and perpendicular magnetic anisotropy (PMA) are needed for future spin-orbitronic device implementations. Here, we show an effect of atomic-scale surface modulation on the magnetic properties and IDMI in ultrathin films composed of *5d* heavy metal/ferromagnet/*4d(5d)* heavy metal or oxide interfaces, such as Pt/CoFeSiB/Ru, Pt/CoFeSiB/Ta, and Pt/CoFeSiB/MgO. The maximum IDMI value corresponds to the correlated roughness of the bottom and top interfaces of the ferromagnetic layer. The proposed approach for significant enhancement of PMA and IDMI through the interface roughness engineering at the atomic scale offers a powerful tool for the development of the spin-orbitronic devices with the precise and reliable controllability of their functionality.

A direct contact between a ferromagnetic metal (FM) and a heavy metal (HM) promotes fascinating spin-orbit coupling (SOC) driven phenomena including perpendicular magnetic anisotropy (PMA)[1], antisymmetric exchange between neighboring spins of *3d* metals (namely, Dzyaloshinskii–Moriya interaction[2, 3, 4, 5]), chiral damping of magnetic domain walls[6], topological spin textures (skyrmions[7, 8], skyrmioniums[9], bimerons[10]), spin Hall effect[11], spin-orbit torques (SOT)[12, 13] and other spin-related effects[14, 15]. The practical applications of these phenomena includes an intensive development of field-free spin current-induced magnetization switching[16, 17], skyrmion racetrack memory[18, 19], SOT nano-oscillators[20], neuromorphic[21, 22] and logic[23] devices.

Recently, it has been shown that the interfacial Dzyaloshinskii–Moriya interaction (IDMI)[24] depends not only on SOC and the lack of the structural inversion symmetry but also on the degree of *3d-5d(4d)* orbital hybridization around the Fermi level[25]. Most of the theoretical studies presenting *ab initio* calculations of the spin-orbit effects assume the atomically smooth interfaces and, consequently, that the IDMI is a constant throughout the interface[2, 26]. Nevertheless, realistic systems have the layer roughness and intermixing on the interfaces at the atomic length scale, causing fluctuations of interaction between the localized spins and heavy metal atoms with the high SOC[27]. As a result, the fluctuating SOC affects the IDMI, surface magnetic anisotropy and, consequently, dynamics of existing chiral spin textures. For the nucleation and stabilization of a spin texture like a skyrmion, the values of the IDMI larger than their critical values for Néel domain walls formation are required[28].

To strengthen the IDMI, the structural inversion symmetry of a layered system has to be broken and the interface quality has to tend to an ideal surface without any disorder[27]. However, even for perfect epitaxially grown layers, a local (at atomic level) variation of IDMI occurs due to the lattice strains[29]. If one induces the opposite signs of IDMI at the top and bottom interfaces of an FM layer using different heavy metals, it is possible to significantly increase this interaction due to the additive effect[30]. One of the promising solutions is to achieve the correlated roughness of both top and bottom interfaces, which cancels the FM layer thickness variation and, consequently, stabilizes the IDMI.

The IDMI strength is extremely sensitive to the thickness of a ferromagnetic layer ($t_{FM}$)[31]. The interface degradation at small $t_{FM}$ stimulates the nonlinear behavior of the IDMI on $1/t_{FM}$. Recently, an influence of the interface quality, considering the surface roughness and atomic intermixing, on the IDMI has been studied in symmetric Pt/Co/Pt systems[32]. Since identical interface qualities or ideally smooth interfaces lead to the vanishing of the IDMI in symmetric systems, such as Pt/Co/Pt and Ru/Co/Ru, an additional treatment could be used (e.g., deposition of films at high temperatures[32, 33] or introduction of an ultrathin intermediate layer of a heavy metal[34]) to induce an interface quality asymmetry. If to assume that in the layered systems with the broken structural inversion-symmetry [25, 30] the quality parameter (it can be determined as the difference between the roughness parameters, such as amplitude and period, of top and bottom interfaces of an FM layer) goes to zero, the IDMI will be maximized. The correlated roughness of both interfaces meets this assumption resulting in the enhancement of PMA as demonstrated theoretically[35].

In this study, we artificially introduce the interface modulation using the atomically smooth Si (111)/Cu surface with the epitaxially grown Pd seed layer of a various nominal thickness ($t_{Pd}$) ranging from 0 to 12.6 nm, as shown in Fig. 1. One of the advantages of the epitaxial Pd seed layer independently of its thickness in the absence of crystal grains affecting the degree of crystallinity and the interface quality of the consequently grown layer of Pt. The root-mean-square (rms) roughness ($R_q$) depends on $t_{Pd}$ and varies from 0.15 to 1.0 nm, which corresponds to the atomic-scale modulation of the surface. First, the multilayer structures with the periodic sub-nm interface modulation were fabricated by a continuous hybrid growth process using a sequence of the molecular beam epitaxy (MBE) for the formation of a Pd buffer layer with the desirable nanoscale surface pattern and the magnetron sputtering for deposition of an HM$_1$/FM/HM$_2$(oxide) layered system with the enhanced PMA and IDMI functionality. As a result, we prepared three series of samples, where each of them consists of the 1.5-nm-thick ferromagnetic CoFeSiB layer asymmetrically sandwiched between one *5d* HM (Pt) and MgO layer, namely the MgO-series, two *5d* HM (Pt and Ta), namely the Ta-series, and one *5d* HM (Pt) from the bottom and *4d* HM (Ru) from the top, namely the Ru-series. The layering scheme of the experimental samples is shown in Supplementary Fig. S1. We report on the direct dependence between the quality of interfaces and the IDMI magnitude. We find that the morphological correlation of the top and bottom interfaces at $t_{Pd}$ = 10.35 nm results in the peak values of the IDMI for the three series of samples.

The Pd island growth leads to the significant increase of $R_q$ as well as the average amplitude ($R_a$) and period ($P_a$) of the Pd surface roughness (see Fig. 1(g)). The persistent increase of the Pd surface roughness is caused by three different growth modes revealed by scanning tunneling microscopy (STM) (the corresponding images are shown in Figs.1(a-d)), indicated in Fig. 1(g), and supported by reflective high energy electron diffraction (RHEED)[36, 37]: (i) Pd grows in 2D mode from the beginning and up to 0.6 nm (Fig. 1(a)), (ii) the layer-by-layer growth mode takes place from 0.6 to 2.9 nm (Fig. 1(b)), and (iii) 3D island growth occurs at $t_{Pd}$ larger than 2.9 nm (Figs. 1(c),(d)). As a result, we fabricated the epitaxial samples with periodically modulated Pd surfaces having the average amplitude and period of an isotropic long-wave roughness ranging from 0.2 to 1.5 nm and 0.75 to 57.0 nm, respectively.

The RHEED pattern of the Pd seed layer with $t_{Pd}$ = 1.125 nm has stripes, and spots are almost not visible, which indicates the reflection diffraction (see insets in Fig. 1(b)). If there was the transmission diffraction, one would see a two-dimensional ordered lattice without curvature, which is formed when the three-dimensional inverse lattice is crossed by the Ewald sphere (it can be approximated by a plane due to its large radius). However, as the thickness of the Pd layer increases, spots appear on the strips as a consequence of the transmission diffraction, which is manifested with a developed surface topography proved by probe methods. It can also be noted that the RHEED background level for the Pd layer with $t_{Pd}$ = 1.125 nm is slightly higher than for other thicknesses, which indicates an increased content of defects at the initial stages of growth.

The background did not change in the other samples (Fig. 1(c,d)) meaning that the quality of the Pd crystal structure remains constant.

The following magnetron sputtering allowed us to grow three series of samples with controllable roughness of each interface as shown in Fig.1(h-j). The roughness data for the Pd and Ta layers were defined experimentally by the STM and AFM, respectively. The layer roughness of the top interfaces of Pt, CoFeSiB, MgO, and Ru was determined by fitting the X-ray reflectivity (XRR) data. We recorded the XRR spectra for incident X-ray beam angles ($\theta$) ranging from 0° to 6°. An example of an XRR spectrum is shown in Supplementary Fig. S2. After performing the fitting procedure in GlobalFit software, we found the best agreement between the experimental and calculated reflexivity curves. As seen in Fig,1(h-j) the $R_q$ values of Pd, Pt, CoFeSiB and the capping layer converge to a point at $t_{Pd}$=10.35 nm for all three series indication the correlation of interfaces.

The analysis of the island profiles (Figs. 1(e,f)) and cross-sectional high-angle annular dark-field scanning transmission electron microscopy (HAADF-STEM) and high-resolution transmission electron microscopy (HRTEM) images (Fig. 2(a)) enabled the identification of each layer and its crystal structure. The highly sensitive to atomic number contrast HAADF-STEM imaging allowed us to distinguish different layers and to determine their thicknesses, which agreed well with the designated growth parameters. The HRTEM imaging in combination with 2D fast Fourier transform (2D FFT) analysis confirmed the epitaxial growth of *fcc (111)*-Pd, Fig.2(b). The observed short-wave roughness induced by the atomic steps was smoothened out by the following magnetron sputtering of 2-nm thick Pt layer. As the lattice mismatch between Pd and Pt was in the range from 1 to 8% depending on the Pd lattice parameter[37], the Pt layer had the crystal structure of Pd: it grew with the same *fcc (111)* texture and repeated the Pd surface modulation, as shown in Figs. 2(a,c). The formed Pd(111)/Pt(111) stack can be considered as a coherent system or a superlattice, because these two metals have the same crystal phase and a small lattice mismatch. Surprisingly, CoFeSiB, which usually has an amorphous structure[38], possessed crystallinity with the *fcc*-like lattice and repeated the surface morphology of the Pd/Pt stack demonstrating practically quasi-epitaxial growth. The following layers of Ru and partially Ta had a pronounced *fcc (111)* crystalline structure as indicated on the 2D FFT patterns in Fig. 2(a). The electron energy loss spectroscopy (EELS) elemental profile is added on the left panel of Fig.2(a) to show the distribution of Pd, Co and Ru atoms in the layers. The upper part of the Ta layer was oxidized for a depth of 3.5 nm from the top. The secondary-ion mass spectroscopy (SIMS) (Fig. 2(d)) and energy-dispersive X-ray spectroscopy (EDX) (Fig. 2(e-g)) proved the intermixing on interfaces and boron diffusion. The Pd layer-induced surface curvature of the correlated interfaces is visible in the magnified HAADF-STEM image of Si(111)/Cu(2.1)/Pd(12.6)/Pt(2)/CoFeSiB(1.5)/Ta sample, Fig.2(h).

It is noteworthy that the samples from the Ta-series sputtered on the Pd layer with thicknesses ranging from 1.125 to 12.6 nm had the four-fold out-of-plane anisotropy with the easy axes tilted at about 45° (225°) to the sample plane, see Supplementary Fig. S3. Samples from the Ru-series had the in-plane magnetic anisotropy only with the disorientation of the easy crystallographic axes within the range from -30° to 30° relative to the sample plane, see Supplementary Fig. S4. Excluding only the very thin Pd layer, all as-deposited samples from the MgO-series had PMA, see Supplementary Fig. S5. These results were supported by the values of the effective magnetic anisotropy $K_{eff}$, shown in Fig. 3(a). The anisotropy energy of the magnetic system can be described as $E(\theta) = -2\pi (M_s^{eff})^2 sin^2\theta + K_1 sin^2\theta + K_2 sin^4\theta$, where $\theta$ is the polar angle of magnetization, $M_s^{eff}$ is the effective saturation magnetization of an FM layer with the effective thickness $t_{eff}$, and $K_1$ and $K_2$ are the first-order and second-order anisotropy constants, respectively. The first-order effective magnetic anisotropy is defined as $K_{eff} = K_1 - 2\pi (M_s^{eff})^2$, where $K_1 = \frac{2K_s}{t_{eff}}$, $K_s$ is the surface anisotropy energy. Samples with $t_{Pd}=$ 10.35 and 12.6 nm from the Ta-series possessed the easy-cone state[39, 40], the formation of which requires the implementation of the following conditions: $K_{eff}<0$, $K_2>0$, and $K_2 > -\frac{1}{2} K_{eff}$. Since we know the cone state angle $\theta=45°$, the $K_2$ value can be calculated using the following relation: $sin\theta = \sqrt{\frac{-K_{eff}}{2K_2}}$ with the solution $K_2 = -K_{eff}$. The $K_{eff}$ values were calculated by fitting the Brillouin light scattering (BLS) spectra (which are discussed below in the paper), providing that the values of $M_s^{eff}$ were defined from the linear slopes of $M_s t_{FM} = f(t_{FM})$ dependences[41], where $t_{FM}$ is the nominal thickness of the FM layer, as shown in Fig. 3(a).

The observed magnetic anisotropy has two main contributions: one arising from spin-orbit (SO) effects and resulting in PMA, and the other coming from long-range dipolar interactions in the presence of randomly periodic magnetic surfaces of high curvature. The SO induced anisotropy may be enhanced by an increased electron scattering when the interface becomes rougher, which is consistent with the $t_{Pd}$ dependence shown in Fig. 3. In the case of magnetic anisotropy due to dipolar interactions, the magnetic interface curvature corresponding to the observed samples should lead to an easy-cone anisotropy[19], which is strikingly confirmed by the above-mentioned measurements. More specifically, the approximation of surface morphology for samples with $t_{Pd}$ = 3.38 and 12.6 nm gave a sinusoidal function with the parameters (see Supplementary Section 4, Fig. S6), which, as calculated according to the method developed in Ref. 19, lead to the easy-cone anisotropy. As shown in Fig. 3(b), the main contribution to PMA of the MgO- and Ta-series brought the surface anisotropy, while, in the Ru-series, the surface contribution was not enough to compete with the demagnetizing energy.

The direct measurement of the IDMI was performed by BLS spectroscopy based on the IDMI-driven asymmetric dispersion shift of long-wavelength thermal spin waves in the Damon-Eshbach surface mode[42, 43]. The effective IDMI energy density ($D_{eff}$) was found from the direct measurements of the frequency shift ($\Delta f$) between Stokes and anti-Stokes spin-wave propagation regimes, $f_s$, and $f_{as}$, respectively:

$$\Delta f = f_s - f_{as} = 2\gamma D_{eff} k / \pi M_s, \qquad (1)$$

where $k$ is the magnon wavevector and $\gamma=176$ GHz/T is the gyromagnetic ratio for CoFeB[44]. The effective saturation magnetization $M_s$ was measured to be 935, 953 and 704 emu/cm$^3$ for MgO-, Ta- and Ru-series, respectively, which is in a good agreement with the experimental data. Examples of the normalized Stokes and anti-Stokes spectra for three series of samples can be seen in Fig. 4(a-c). The corresponding field and wavevector dependences of peak frequency shifts are shown in Fig. 4(d-f). In the case of surface spin-wave modes (the Damon-Eshbach geometry) in ultrathin magnetic films, spin waves are localized on the top or bottom interfaces depending on the direction of the wave vector $k$. If these interfaces have different surface anisotropies ($K_{S1}$ and $K_{S2}$ are surface anisotropies of the bottom and top interfaces, respectively), then the spin waves are affected by different effective fields on the top and bottom interfaces, which in turn lead to different frequencies of the Stokes and anti-Stokes peaks. Thus, the difference in surface anisotropy can contribute to $\Delta f$. To separate contributions from IDMI and $K_S$, it is necessary to evaluate the influence of $K_{S1}$ and $K_{S2}$ on $\Delta f$. To do this, we used an analytical model of dipole-exchange spin waves[45], in the framework of which the frequency shift due to $K_S$ can be calculated as:

$$\Delta f_k = \frac{8\gamma}{\pi^2}\frac{K_{S1}-K_{S2}}{M_S}\frac{k}{1+\frac{l_{ex}^2\pi^2}{t_{FM}^2}}, \qquad (2)$$

where $l_{ex} = \sqrt{\frac{2A}{4\pi M_S^2}}$ is the exchange length. This equation is an approximation for thin ferromagnetic films ($kt_{FM}\ll 1$), as in the case of our experiments. The calculation of the surface anisotropy effect gives values of $\Delta f < 50$ MHz (this value is comparable with the BLS measurement error), while the experimentally observed $\Delta f$ values can achieve 1.6 GHz. This fact allows concluding that IDMI is the main source of the frequency shift in our systems. Moreover, the observed linear dependence of $\Delta f$ on the wave vector $k$ proves that the frequency shift is due to the anti-symmetrical exchange inducing by the IDMI[31, 45].

As the reference samples for the MgO-, Ta- and Ru-series, we used Si/SiO$_2$/Pt(2)/CoFeSiB(1.5)/MgO(1)/Ta(5), Si/SiO$_2$/Pt(2)/CoFeSiB(1.5)/Ta(5), and Si/SiO$_2$/Pt(2)/CoFeSiB(1.5)/Ru(3)/Ta(5) films, respectively. All these as-deposited films had the in-plane anisotropy only and lowest surface roughness. Recently, it has been shown that Pt, Ta, and Ru have the negative sign of the IDMI at a bottom interface leading to the partial cancellation of the effective IDMI ($D_{eff}$) if a ferromagnet layer is sandwiched between them[46, 47]. We defined by BLS that these three reference samples have negative effective values of the IDMI: $D_{eff}$ = -0.83, -0.48, and -0.34 erg/cm$^2$, respectively. If we assume that the CoFeSiB/MgO interface had zero IDMI, then the CoFeSiB/Ta and CoFeSiB/Ru interfaces possess the positive IDMI with the approximate values of 0.35 and 0.49 erg/cm$^2$, respectively. This result proves the partial compensation effect of the IDMI induced at the bottom and top interfaces in a Pt ($D_{top}$ <0)/CoFeSiB/Ru or Ta ($D_{bottom}$ >0) structure.

The increasing thickness of the Pd seed layer leads to the significant changes of the IDMI compared to the reference samples, as shown in Fig. 4(g). The sputtering of the corresponding layer sequence on the Si (111)/Cu(2.1)/Pd surface with $t_{Pd}$ = 1.125 nm sharply decreased $D_{eff}$ down to -0.33±0.04 erg/cm$^2$ for all the series of samples. The following increase of $t_{Pd}$ up to 10.35 nm leads to the significant increase of $D_{eff}$, see Fig. 4(g). We excluded an effect of the degree of crystallinity and size of Pt(111) grains emerging in polycrystalline or textured systems, which may change with the Pd seed layer thickness, on the strength of the exchange interactions at the interface,[48] because the grown Pd(111)/Pt(111) stacks have the epitaxial structure without the formation of crystallites as supported by the complex HAADF-STEM, HRTEM, RHEED and X-ray diffraction analyses where the XRD spectra and their descriptions are shown in Supplementary Fig. S7. Thus, our findings show that the small variation of the periodic surface roughness with the amplitude less than 1.0 nm provided the Pt/CoFeSiB interface modulation at the atomic scale, which was enough for a drastic enhancement of the IDMI value. In the Ru-series at $t_{Pd}$ > 10.35 nm, the effective IDMI rapidly decreased down to -0.4 erg/cm$^2$, while it stayed constant for the MgO- and Ta-series. The dependence of $D_{eff}$ on the amplitude of roughness $R_a$ is demonstrated in Fig. 4(h).

The IDMI enhancement mainly comes from the increased electron scattering at the interfaces in the presence of strong spin-orbit interaction, in the same way spin-orbit torques are enhanced[46]. Based on previous theoretical work, it is expected that disorder may further increase the Dzyaloshinskii-Moriya interaction strength since it increases the scattering of electrons. The latter, in the presence of an emergent magnetic field due to the spin-orbit interaction, gives rise to spin-dependent scattering, thus leading to additional spin accumulation at the interface with the ferromagnet. This accumulation of spins, in turn, may exert a torque on the magnetization[49]. Therefore, at highly disordered interfaces as those studied here, the effects due to spin-orbit coupling, such as IDMI, are expected to be enhanced. The dependence of the IDMI and the spin Hall effect on disorder strength may be probed based on the nonequilibrium Green function Keldysh formalism[50]. This scattering is increased when the interfacial surface becomes less smooth, and especially when the intermixing is present. The overall increase of the effective interfacial surface contributes to this enhancement as well.

Thus, we showed the correlation between the Pd layer thickness (or surface roughness) and the magnitude of the effective IDMI. However, to find a direct correlation, one has to analyze the roughness and intermixing on Pt/CoFeSiB and CoFeSiB/MgO(Ta, Ru) interfaces. A cross-sectional HRTEM observation does not allow to directly inspect an interfacial intermixing. To qualitatively analyze the layer compositions and possible intermixing, we employed SIMS using $Cs^+$ ions (an example is shown in Fig.2(d)). For all the series of samples, we found that the intermixing on the Pt/CoFeSiB interface did not depend on the Pd layer thickness and kept a constant depth. The intermixing depth of the top CoFeSiB/MgO interface was practically zero, which is in agreement with the previous studies[51,52,53]. On the contrary, the top CoFeSiB/Ta and CoFeSiB/Ru interfaces demonstrated an increased intermixing depth with an increasing $t_{Pd}$. The most reliable quantitative method for this estimation is XRR, which enabled us to extract the $R_q$ value (Fig.1(h-j)) and average intermixing depth ($I_d$) for each interface. This approach gives a set of parameters of multilayer structure, including surface/interface roughness, layer thickness, and density variation due to the intermixing[54].

To analyze the quality of interfaces, we introduced the quality factor $\Delta\sigma/\sigma$, where $\Delta\sigma$ is the difference between the roughness of the top and bottom interfaces of the CoFeSiB layer ($R_q^{top} - R_q^{bottom}$) and $\sigma$ is the sum of roughness ($R_q^{top} + R_q^{bottom}$). The dependence of $R_q$ for top interfaces of all the layers with the Pd thickness is represented in Fig. 1(h-j). The main trend was the increase of the interface roughness for all the layers with the increasing $t_{Pd}$. Noteworthy, the initial $R_q$ for CoFeSiB was higher than for the other layers. It could be related to the initially amorphous nature of this ferromagnetic material. With an increase of $t_{Pd}$ up to 10.35 nm, the roughness of the CoFeSiB layer approached to $R_q$ of the underneath Pt layer, which corresponds to the roughness correlation of the top and bottom interfaces. The experimental data revealed that the Ta capping layer smoothed out the surface roughness. Based on these data, we plotted the dependence $\Delta\sigma/\sigma = f(t_{Pd})$ for the Ru- and Ta-series, as shown in Fig. 4(i). One can see the correlation between the interface quality factor and the surface DMI constant ($D_s$) for both series. The $D_s$ constant is the thickness-independent indicator of the IDMI strength and can be defined as $D_s = D_{eff} t_{eff}$, where $t_{eff}$ is the effective thickness of a ferromagnetic layer considering the magnetically dead layer (MDL) thickness ($\Delta$) of the bottom and top interfaces ($\Delta = \Delta_{bottom} + \Delta_{top}$). From the XRR analysis, we found that the intermixing depth ($I_d$) on the Pt/CoFeSiB interface was the same for samples from the three series ($I_d$ = 0.24 nm) and it did not depend on $t_{Pd}$. According to our investigation and studies conducted by different groups, $\Delta_{bottom}$ for a Pt/FM interface is zero or even negative due to the effect of the proximity induced magnetization[55]. However, the top CoFeSiB/MgO, CoFeSiB/Ta, and CoFeSiB/Ru interfaces had different values of $I_d$: it was zero for CoFeSiB/MgO, the intermixing depth changed from 0.22 to 0.47 nm with an increase of $t_{Pd}$ for the CoFeSiB/Ta, and for CoFeSiB/Ru, the latter parameter, $I_d$, increased with an increase of $t_{Pd}$ (or Pd roughness) from 0.15 nm for $t_{Pd}$ = 1.125 nm up to 0.33 nm for $t_{Pd}$ = 12.6 nm. The largest $I_d$ for the CoFeSiB/Ta interface is owing to the more intense interdiffusion between layers. In addition, the observed fact can be explained in terms of interfacial enthalpy, which is a source of interdiffusion, and its values are various for different materials[51,56]. Since the intermixing is the main reason for the MDL formation[57,58], we suggest that $I_d$ coincided with $\Delta_{top}$ for CoFeSiB/Ta and CoFeSiB/Ru interfaces. The defined MDL depths for Ta and Ru are in a good agreement with the data found for the CoFeB/Ta(Ru) interfaces[53,59,60]. Finally, one can find that $D_s = D_{eff}(t_{FM} - \Delta_{top})$. This formula was used to calculate $D_s$ as subject to $t_{Pd}$ shown in Fig. 4(i).

With a small Pd roughness, the magnitude of $\Delta\sigma/\sigma$ is a maximum, which corresponds to the highly uncorrelated top and bottom interfaces of the CoFeSiB layer, leading to the relatively small $D_s$ values. The flowing tendency of $\Delta\sigma/\sigma$ with increasing $t_{Pd}$ reflects the growing correlation of the interfaces, causing a significant increase of $D_s$ about twice for the three series. The same increase of $D_s$ continuing up to $t_{Pd} = 10.35$ nm suggests the same nature of this effect in all the series of samples. The further increase of $t_{Pd}$ and, consequently, its $R_q$, whose value compares with the CoFeSiB thickness, is accomplished by the decrease of $D_s$ for the Ru- and Ta-series. The maximum value of $D_s = -1.1 \times 10^{-7}$ erg/cm was found for the Pt/CoFeSiB/MgO system and it can be compared with other Co-based systems such as Pt/Co$_2$FeAl$_{0.5}$Si$_{0.5}$/MgO ($D_s = -0.42 \times 10^{-7}$ erg/cm, measured by BLS)[61], Pt/Co/AlO$_x$ ($D_s = -1.7 \times 10^{-7}$ erg/cm[62] and $1.4 \times 10^{-7}$ erg/cm[63], measured by BLS), Pt/CoFeB/MgO and Pt/CoFe/MgO ($D_s = 0.8 \times 10^{-7}$ erg/cm[64] and $-1.27 \times 10^{-7}$ erg/cm[65], measured by BLS), Pt/Co/MgO ($D_s = 2.17 \times 10^{-7}$ erg/cm, measured by BLS)[26], Pt/Co/Ir/Pt ($D_s = 0.84 \times 10^{-7}$ erg/cm, measured by the asymmetric magnetic domain growth)[34], and Ta/Pt/Co/Ir ($D_s = 1.6 \times 10^{-7}$ erg/cm and $2.2 \times 10^{-7}$ erg/cm, measured by BLS and the asymmetric hysteresis, respectively)[66]. It has to be noted that, in most of the papers, the magnetically dead layer depth is not taken into account for the calculation of $D_s$, resulting in the overestimated values. As shown, the magnitudes of the IDMI are particularly similar for the same systems. Meanwhile, the IDMI signs, even for the same IDMI measurement method, are opposite. Such discrepancy for the BLS results may be a product of the incorrect definition of the Stokes and anti-Stokes peaks relative to the applied in-plane magnetic field. A case where a Stokes peak has a larger intensity than an anti-Stokes one has to be considered as measured at the positive magnetic field and vice versa, see Fig. 4(a-c). Then, if the Stokes frequency ($f_S$) is smaller (larger) than anti-Stokes ($f_{AS}$), the IDMI value will be negative (positive) since $D_s \sim f_S$-$f_{AS}$.[42] Based on this fact, our analysis of the works representing the IDMI data measured by BLS revealed that, in contrast to the shown data[26, 31, 64, 65, 66], the sign of $D_s$ for Pt/Co-based systems is negative, promoting the formation of left-handed cycloidal spin structures[2].

The found effect of the periodic surface modulation on the IDMI can be explained within a model of correlated roughness of interfaces, where the amplitude and period of roughness are the same for both the top and bottom interfaces of the CoFeSiB layer and the parameter $\Delta\sigma/\sigma$ goes to zero. In this case, the thickness of CoFeSiB remains unchanged giving, as a result, near the maximum IDMI value, which has to be similar for ideally smooth interfaces. For the uncorrelated roughness regime, the period of roughness is the same, but the roughness amplitude for the top and bottom interfaces is different ($R_q^{top} > R_q^{bottom}$). In this case, $\Delta\sigma/\sigma >> 0$, which means there are local variations of the CoFeSiB layer thickness and, consequently, local variations of the IDMI.

In this article, the roughness-dependent enhancement of the IDMI has been successfully demonstrated for the first time. We have shown that the correlated roughness of the top and bottom interfaces can increase the IDMI values by up to 2.5 times, with the maximum observed value being $D_s = -1.1 \times 10^{-7}$ erg/cm, which is the largest known IDMI for CoFeB-based systems. The main reason for this enhancement is the intermixing at the bottom and top interfaces as well as the correlated interface-roughness variations, which can both affect electronic transport across the interface and, as a result, the degree of the electron scattering. The latter is the main driver of the IDMI and other spin-orbit effects[49].

**Methods**

**Sample preparation.** The Si(111) substrates/Cu(10 ML)/Pd(0–56 ML) with the roughness variation were evaporated in an ultrahigh vacuum complex Omicron Nanotechnology, consisting of a molecular beam epitaxy chamber and an analytical chamber, which were interconnected with each other[37]. Si (111) substrates misoriented towards [11-2] by 0.1° were used. Before being loaded into the chamber, the Si (111) substrates were rinsed with isopropyl alcohol and distilled water. Then, the substrates were heated at 500 °C by indirect heating for 12 h. Immediately before deposition, samples were flash-heated by direct current at 1200 °C three times for 10 s and slowly cooled down to 50 °C. All the metals were evaporated from the high-temperature effusion cells. The thickness of the layers was measured in monolayers (ML): 1 ML Cu corresponds to the thickness of one ideal layer of Cu, which is equal to 2.09 Å; 1 ML of Pd is equal to 2.25 Å. The growth rates of Cu and Pd were 4.3 and 0.75 ML/min, respectively. The temperature of the substrate was 75 °C during both the Cu and Pd depositions. To prevent intermixing of Pd and Si and to initiate the epitaxial growth of *fcc*-Pd(111), a Cu (10 ML) buffer layer was formed on the Si surface before the deposition of the Pd buffer layer. The lattice period of the Pd during growth and its structure was analyzed employing reflection high-energy electron diffraction (RHEED, Staib Instruments). The RHEED measurements were done simultaneously with the deposition of the samples. The topography of the Pd buffer layer was investigated *in situ* using a scanning tunneling microscope manufactured by Omicron Nanotechnology. The Pd surface roughness was defined *ex-situ* on areas of 5×5 µm$^2$ and 2×2 µm$^2$ using an atomic force microscope (AFM, Ntegra Aura, NT-MDT). Immediately after the MBE process, the following three series of samples were deposited on the top of the Pd surface by magnetron sputtering: MgO-series - Pt(2)/CoFeSiB(1.5)/MgO(2)/Ta(5), Ta-series - Pt(2)/CoFeSiB(1.5)/Ta(5), and Ru-series - Pt(2)/CoFeSiB(1.5)/Ru(3)/Ta(5), where the layer's thickness is indicated in nm, as shown in

Supplementary Fig. S1. The ferromagnetic layer has the composition $Co_{70.5}Fe_{4.5}Si_{15}B_{10}$ represented in at.% [38].

**Atomic structure observation.** The scanning transmission electron microscopy (STEM) images and electron energy loss spectra (EELS) were collected using Cs-corrected microscope Titan 80-300 operated at 300 kV and equipped with a Gatan Quantum 966 spectrometer. The convergence and collection angles were 24.9 and 24.7 mrad, respectively. TEM sample preparation was performed in a precision ion-polishing system (PIPS, Gatan model 691) with an ion beam energy of 3.5 keV and milling angle of 6º in double sector milling mode. To prevent a temperature increase, the sample was cooled to -165ºC using liquid nitrogen during milling. To characterize the elemental depth profile, secondary ion mass spectrometry (SIMS, TOF.SIMS 5) using $Cs^+$ ions was utilized.

**Surface and interface roughness measurements.** The surface morphology was studied using STM and AFM (Ntegra Aura, NT-MDT). To analyze the crystal structure and interface quality (roughness, intermixing, and thickness variation), we used the analytical methods including X-ray diffraction (XRD) and X-ray reflectivity (XRR). The study was performed on a SmartLab (Rigaku) X-ray diffractometer at CuKα radiation wavelength (1.54 Å). The simulations of XRR spectra were performed with GlobalFit software.

**Magnetic characterization.** The magnetic properties of the films were investigated through the magneto-optical Kerr effect (NanoMOKE II, Durham Magneto Optics) and by using a vibrating sample magnetometer (7410 VSM, LakeShore). Magnetic force microscopy (Ntegra Aura, NT-MDT was used to observe the magnetic domain structure in an applied magnetic field using low magnetic moment tips. Magnetization reversal processes were studied by a Kerr microscope (Evico Magnetics).

**Acknowledgments**

This research was supported in part by the Future Materials Discovery Program through the National Research Foundation of Korea funded by the Ministry of Science and ICT (No. 2015M3D1A1070465), by the Samsung Electronics' University R&D program, by the Russian Foundation for Basic Research (grant 19-02-00530), by the Russian Ministry of Education and Science under the state task (FZNS-2020-0013), and by Act 211 of the Government of the Russian Federation (No. 02.A03.21.0011).

**Author contributions** A.S.S., A.V.O., O.A.T, and Y.K.K. generated the idea of the surface roughness effect and proposed the project. A.V.D., A.G.K., I.H.C., and Y.J.K. prepared the experimental samples. B.P. and A.Yu.S. measured magnetic properties. A.V.S. and S.A.N. performed BLS spectroscopy. A.V.G. carried out XRR and XRD studies. A.G.K. performed the micromagnetic simulations and analyzed them with the assistance of O.A.T. The interpretation of the experimental data was performed by A.S.S. and O.A.T. with the assistance of A.V.O. A.S.S. analyzed the experimental data and wrote the majority of manuscript with the help of O.A.T. and Y.K.K.

**Competing interests statement** Authors declare that they have no competing financial interests.

**Correspondence** and requests for materials should be addressed to A.S.S. (e-mail: samardak.as@dvfu.ru) or Y.K.K. (e-mail: ykim97@korea.ac.kr).

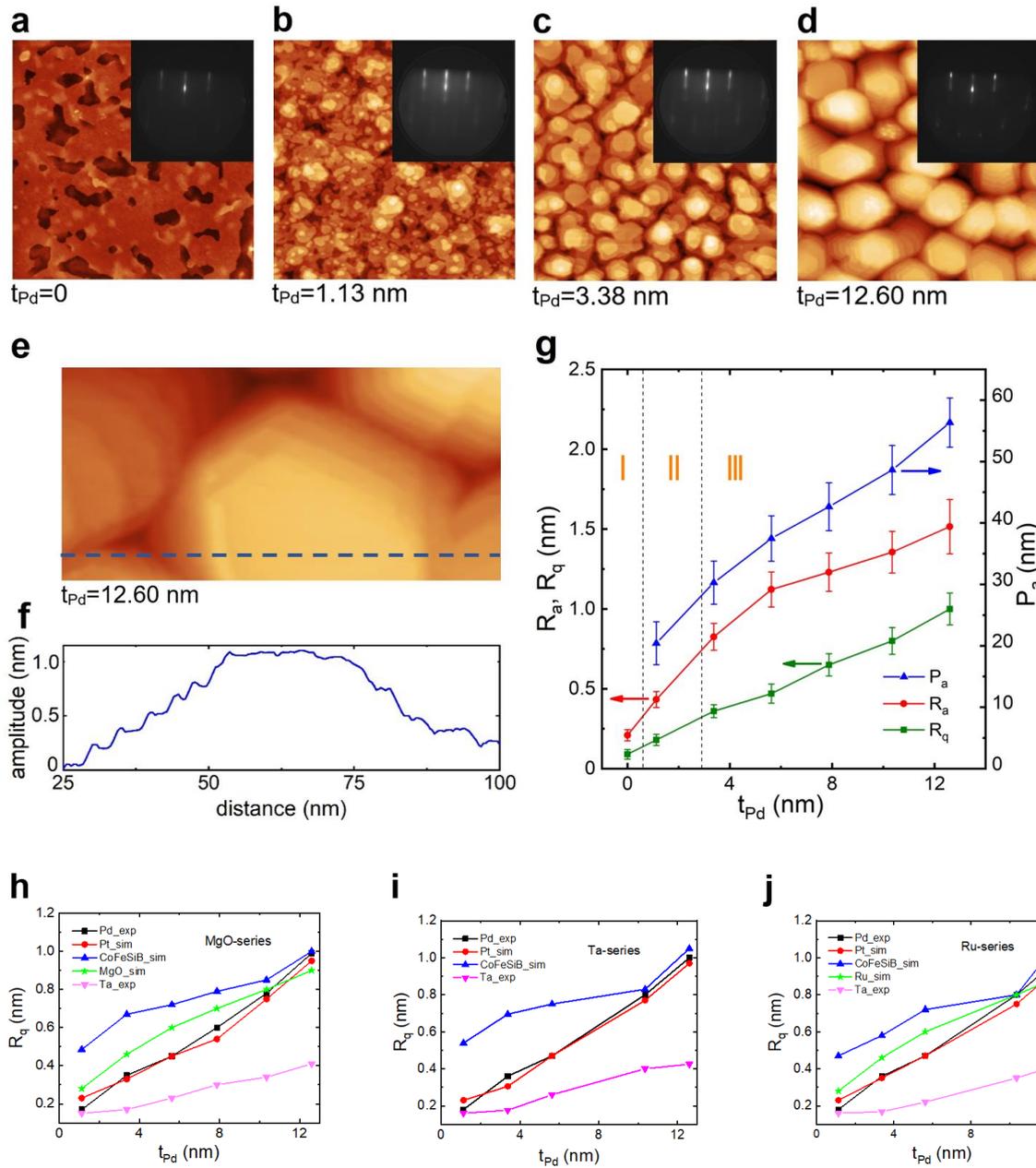

**Fig. 1.** (a-d) The 200×200 nm² STM images of Si(111)/Cu(2.1)/Pd surface of various thicknesses. Insets show RHEED patterns for corresponding samples. (e) Magnified STM image of a Pd island-grown during the 12.6-nm-thick Pd layer deposition and (f) its corresponding *x*-profile. (g) Dependence of the average amplitude of roughness ($R_a$), root-mean-square roughness ($R_q$), and average period of roughness ($P_a$) on the nominal thickness of the Pd buffer layer ($t_{Pd}$). Root-mean-square roughness ($R_q$) of layers for (h) MgO-, (i) Ta-, and (j) Ru-series as a function of $t_{Pd}$. Experimental values were measured by STM and AFM. Simulation values were calculated from the fitting of the experimental XRR spectra.

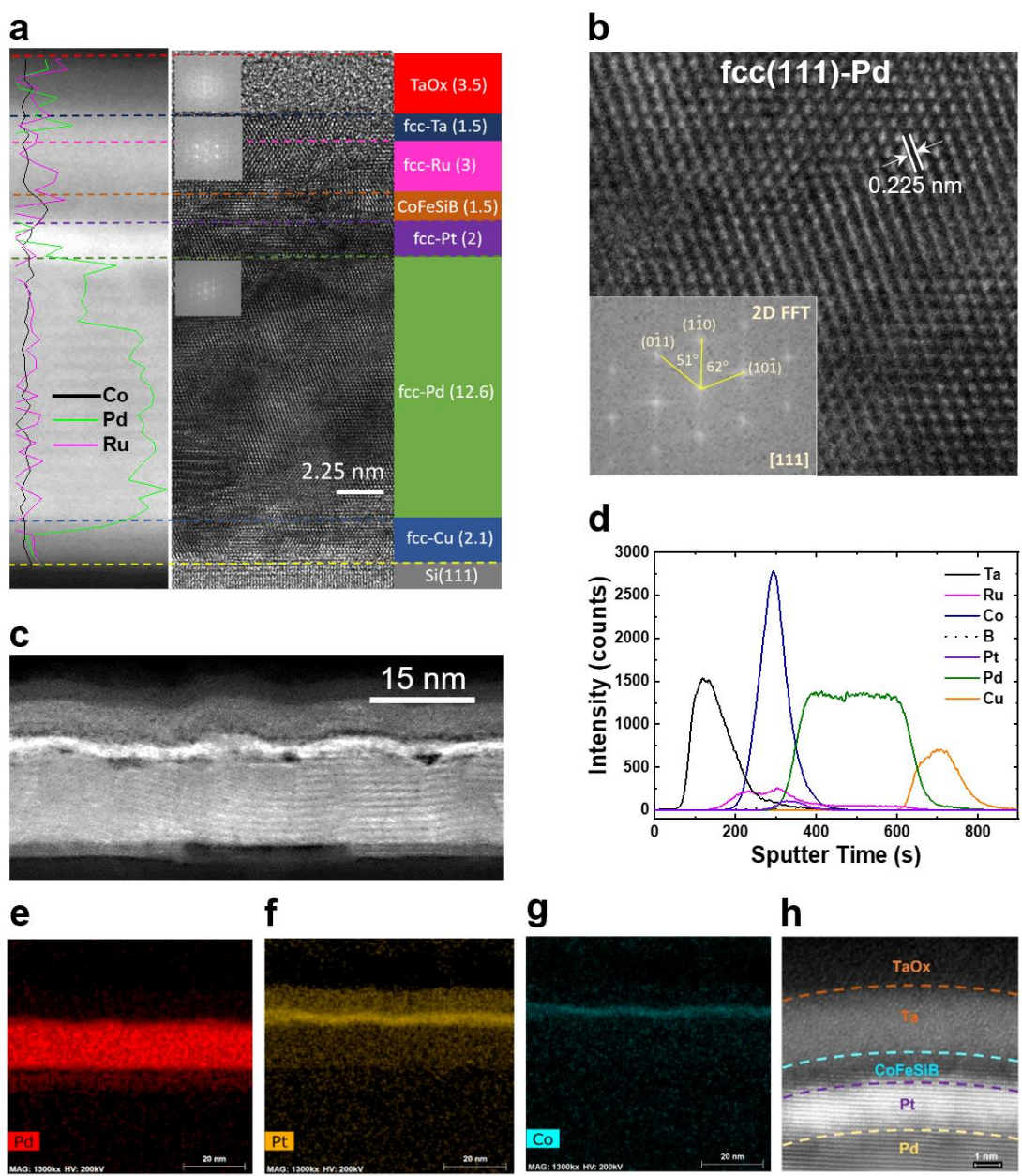

**Fig. 2.** (a) Cross-sectional HAADF-STEM with the EELS elemental analysis (left) and HRTEM (right) images of the Si(111)/Cu(2.1)/Pd(12.6)/Pt(2)/CoFeSiB(1.5)/Ru(3)/Ta(5) sample. Insets in the HRTEM image consist of 2D FFT patterns for corresponding areas. (b) Atomic-resolution imaging of the Pd layer and corresponding 2D FFT pattern supporting *fcc(111)*-Pd phase formation. (c) Large-area cross-sectional HAADF-STEM image enabling to observe the surface modulation. (d) Original SIMS spectrum of the designated element distribution for the presented sample. Note that boron (B) intensity is very low to be recognized in the original ion counts. (e-g) EDS elemental maps for the Si(111)/Cu(2.1)/Pd(12.6)/Pt(2)/CoFeSiB(1.5) sample. (h) The magnified high-resolution HAADF-STEM image with the visible curvature of correlated interfaces. The dashed lines are shown for guidance only.

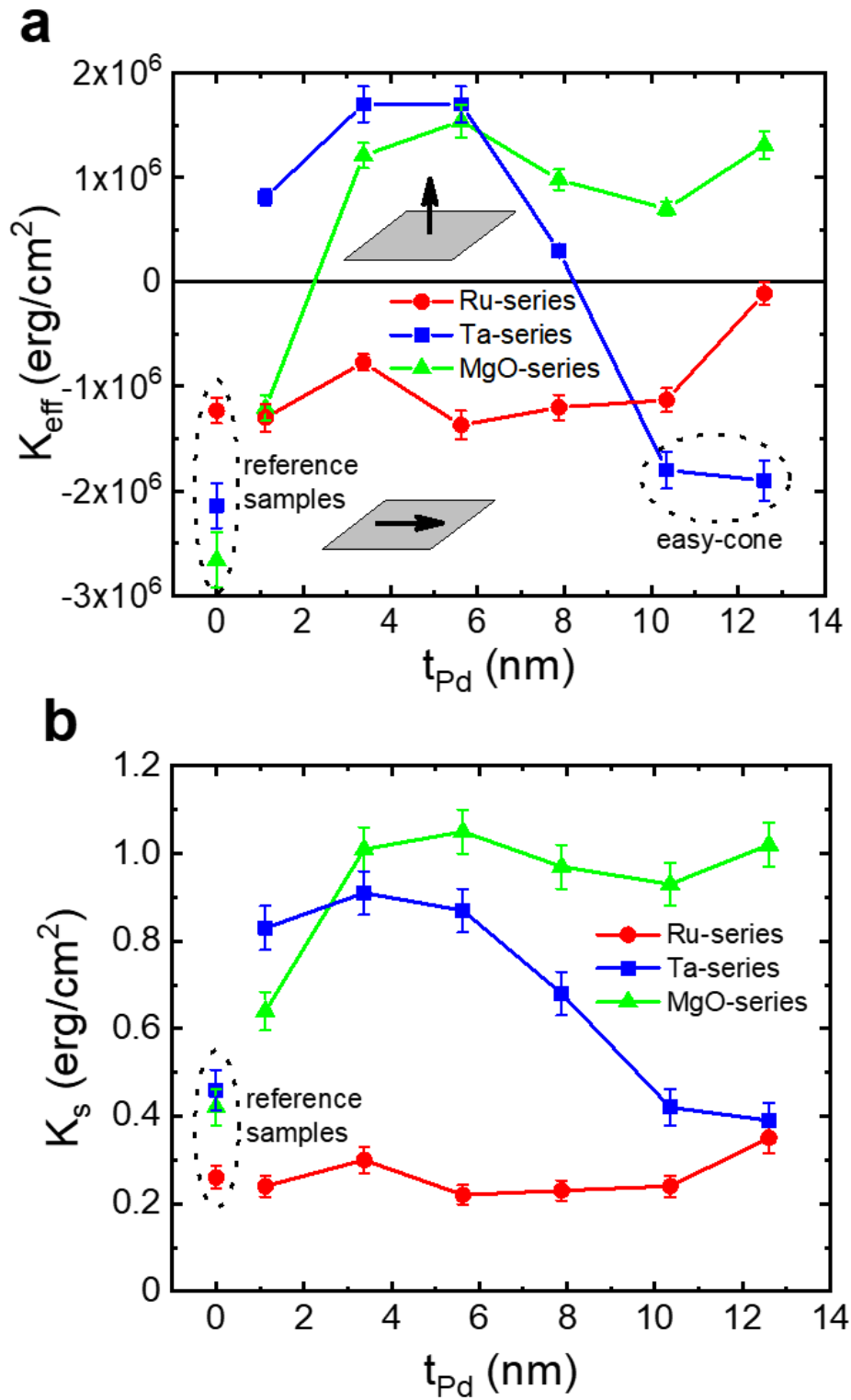

**Fig. 3**. (a) Effective magnetic anisotropy energy and (b) surface anisotropy energy as a function of $t_{Pd}$ for the three series of samples.

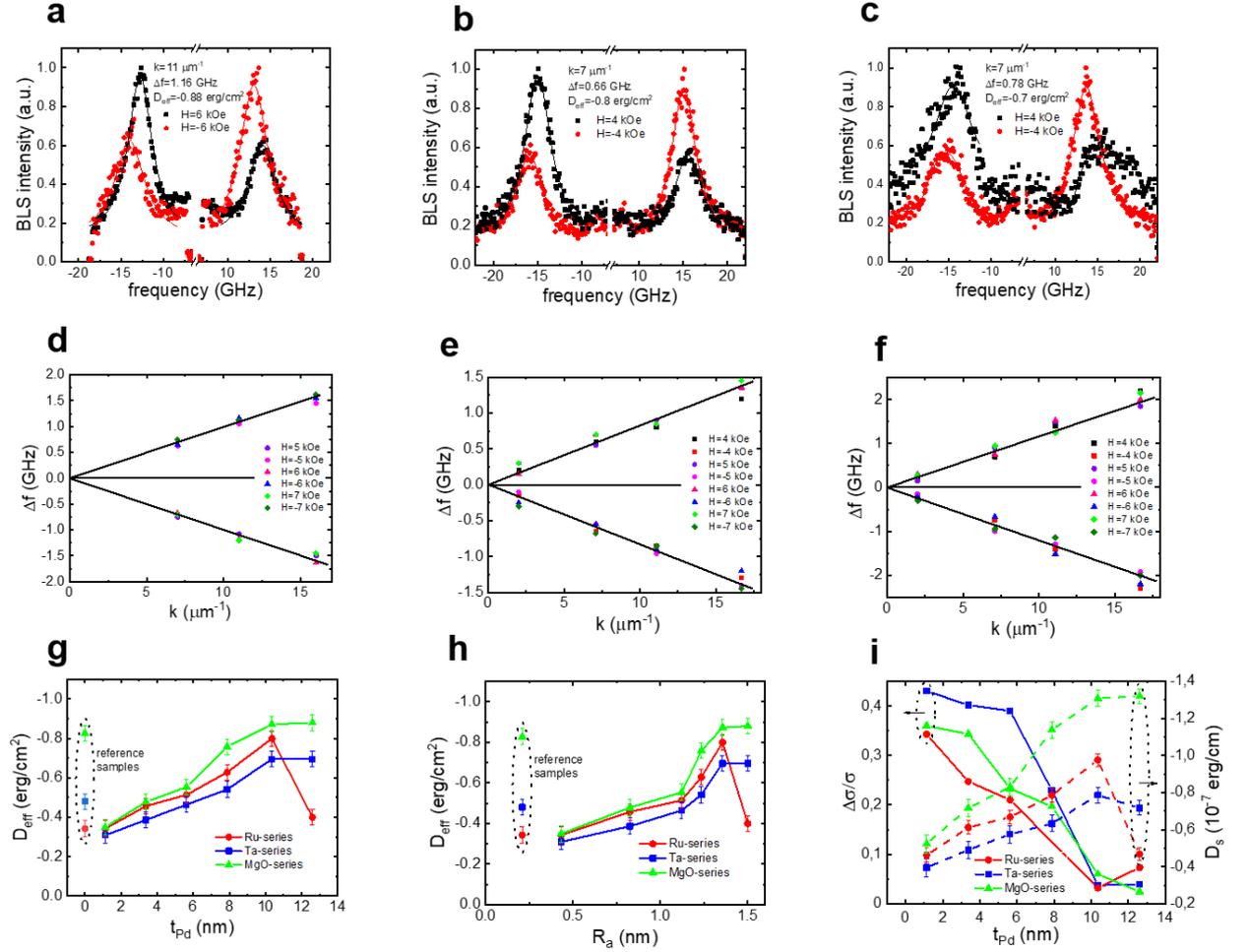

**Fig. 4**. High-resolution BLS spectra and corresponding dependences of peak frequency shifts $\Delta f$ on the wave vector $k$ measured in positive and negative fields for films with $t_{Pd}$ = 10.35 nm taken from the (a, d) MgO-, (b, e) Ta-, and (c, f) Ru-series. The solid lines reflect the fit of the experimental data. Dependences of $D_{eff}$ on the nominal thickness of the Pd layer (g) and on the amplitude of roughness $R_a$ (h) for MgO-, Ta- and Ru-series. (i) The interface quality factor $\Delta\sigma/\sigma$ and surface DMI constant ($D_s$) depending on $t_{Pd}$ for the three series of samples.

# Supplementary Information

# Enhancement of perpendicular magnetic anisotropy and Dzyaloshinskii-Moriya interaction in thin ferromagnetic films by atomic-scale modulation of interfaces

A. S. Samardak,[1,5*] A. V. Davydenko,[1] A. G. Kolesnikov,[1] A. Yu. Samardak,[1] A. G. Kozlov[1], Bappaditya Pal,[1] A. V. Ognev,[1] A. V. Sadovnikov,[2,3] S. A. Nikitov,[2,3] A. V. Gerasimenko,[4] In Ho Cha,[6] Yong Jin Kim,[6] Gyu Won Kim,[6] Oleg A. Tretiakov,[7,8] Young Keun Kim[6*]

[1]*School of Natural Sciences, Far Eastern Federal University, Vladivostok 690950, Russia*

[2]*Laboratory "Metamaterials," Saratov State University, Saratov 410012, Russia*

[3]*Kotel'nikov Institute of Radioengineering and Electronics, Russian Academy of Sciences, Moscow 125009, Russia*

[4]*Institute of Chemistry, Far East Branch, Russian Academy of Sciences, Vladivostok 690022, Russia*

[5]*National Research South Ural State University, Chelyabinsk, Russia*

[6]*Department of Materials Science and Engineering, Korea University, Seoul 02841, Republic of Korea*

[7] *School of Physics, The University of New South Wales, Sydney 2052, Australia*

[8] *National University of Science and Technology ``MISiS'', Moscow 119049, Russia*

*e-mails: samardak.as@dvfu.ru; ykim97@korea.ac.kr

## 1. Experimental samples and structural characterization

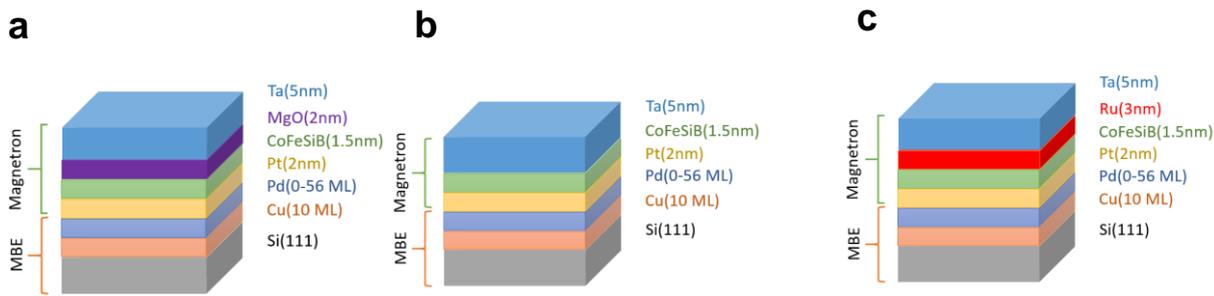

**Fig. S1**. Layer composition of samples from the (a) MgO-, (b) Ta-, and (c) Ru-series.

## 2. Interface characterization

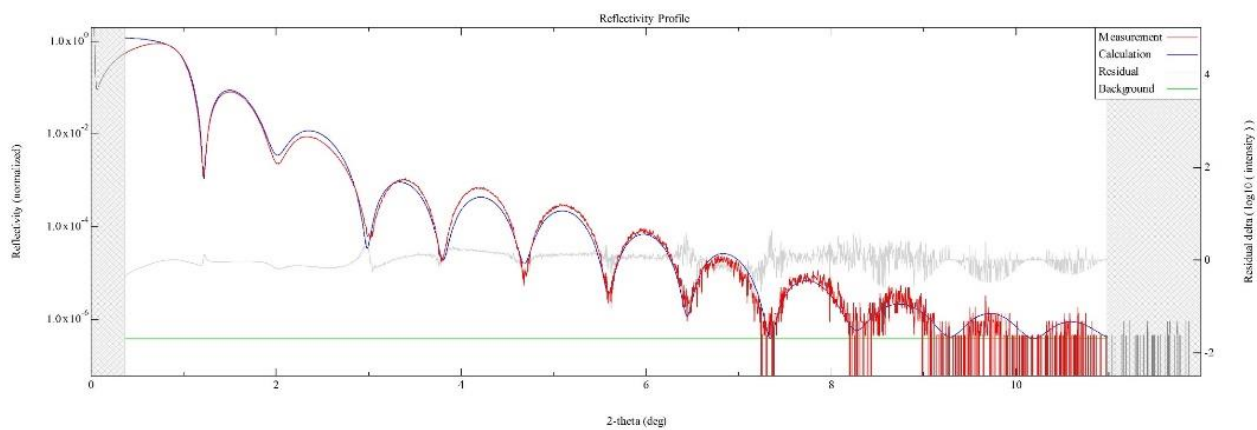

**Fig. S2**. XRR spectrum and calculated reflectivity curve for the sample with the 1.125-nm thick Pd from the Ta-series.

## 3. Magnetic characterization

The polar diagram $M_r/M_s = f(\varphi)$ measured in out-of-plane geometry for the Ta-series revealed the presence of four-fold out-of-plane anisotropy with two easy axes lying at 45° and 135° to the sample plane, as shown in Fig. S3(a). The presence of two canted out-of-plane easy axes can be evidence of "easy-cone" anisotropy[1], which was found in different systems including Co/Pt multilayers[2]. The observed canted magnetization appears due to the competition between perpendicular magnetic anisotropy (PMA), which originates from both the interface and internal magnetic atoms[3], and shape anisotropy (demagnetizing and trying to turn magnetization into the plane) and corresponds to the spin reorientation transition (SRT) from the perpendicular easy axis to the easy plane[4].

The in-plane $M_r/M_s = f(\varphi)$ polar diagram for the Ta-series is shown in Fig. S3(b). It is seen that samples are practically isotropic in the plane. Figures S3(c) and (d) demonstrate the magnetic hysteresis loops for all samples in the Ta-series measured in the magnetic field applied at 45° and 135°, respectively. As shown, the coercive force increases with an increase of the Pd layer thickness. This fact is correlated with the surface roughness fashion since the domain walls have a higher probability of being pinned by magnetostatic fields in a rough film than in a smooth one[5]. The domain walls magnetostatically interact with the stray fields of roughness. As a result, domains nucleate in the larger fields, consequently providing the higher coercivities. Samples are practically isotropic in-plane. The presence of the out-of-plane component of the magnetic anisotropy was confirmed by MOKE measurements, as shown in Fig. S3(e).

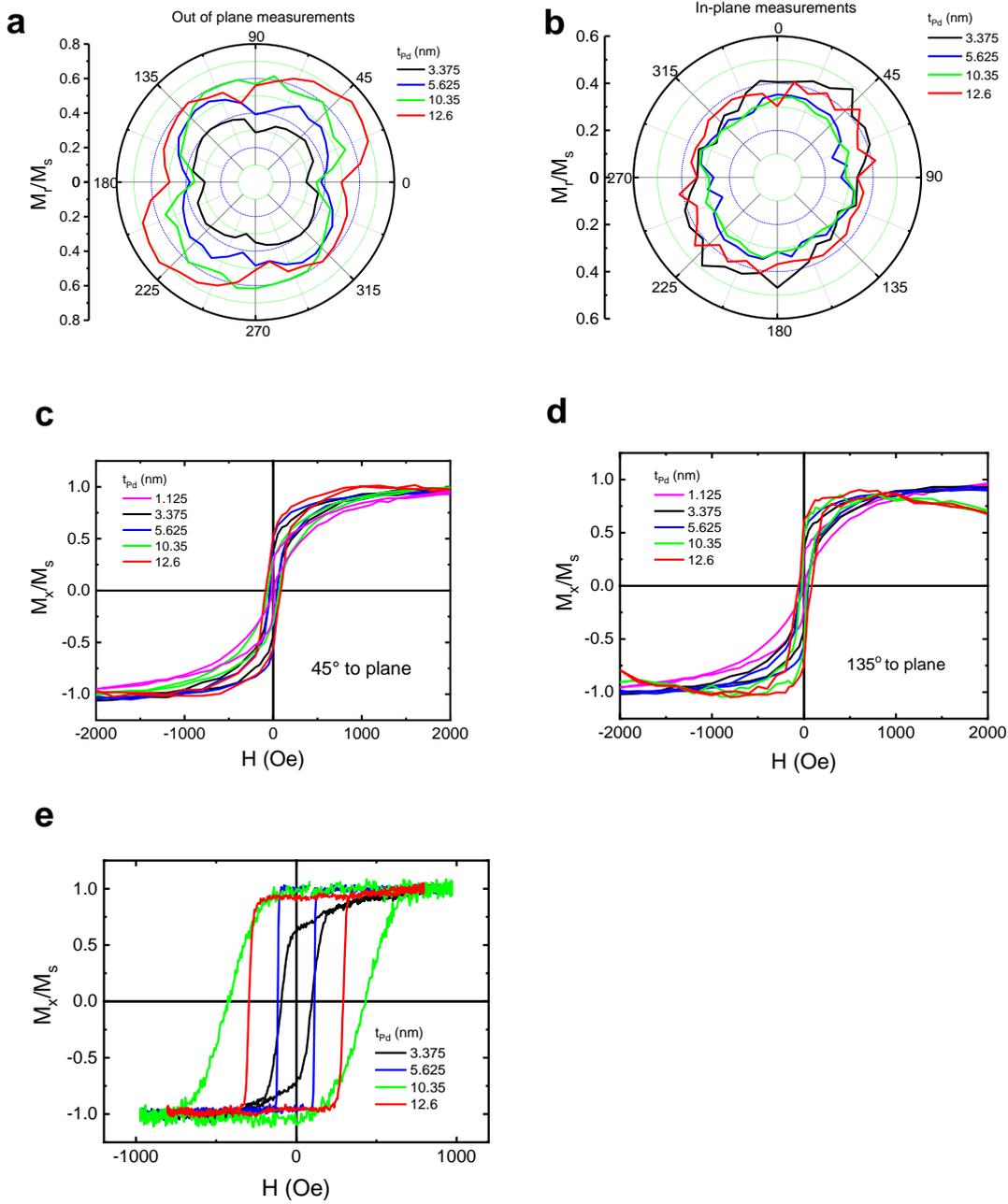

**Fig. S3.** Polar diagrams of $M_r/M_s=f(\varphi)$ for (a) out-of-plane and (b) in-plane measurements of the Ta-series. Hysteresis loops for canted easy magnetization axes for the Ta-series measured in the magnetic field applied at (c) 45° and (d) 135° to the sample plane. (e) Hysteresis loops recorded by MOKE in out-of-plane geometry for Ta-series.

A completely different situation was observed in the Ru-series. The sputtering of the Ru layer on the top interface of CoFeSiB induced easy-plane anisotropy with wide-angle out-of-plane distribution of magnetization, as shown in Fig. S4. The explanation can be the following. The Pt layer, by its *fcc* crystal structure, defines PMA in the CoFeSiB layer[3], but Ru has an *hcp* phase and destroys PMA due to the compressive stress affecting the short-range exchange interaction[6]. The coercivities of this series are independent of the Pd layer thickness.

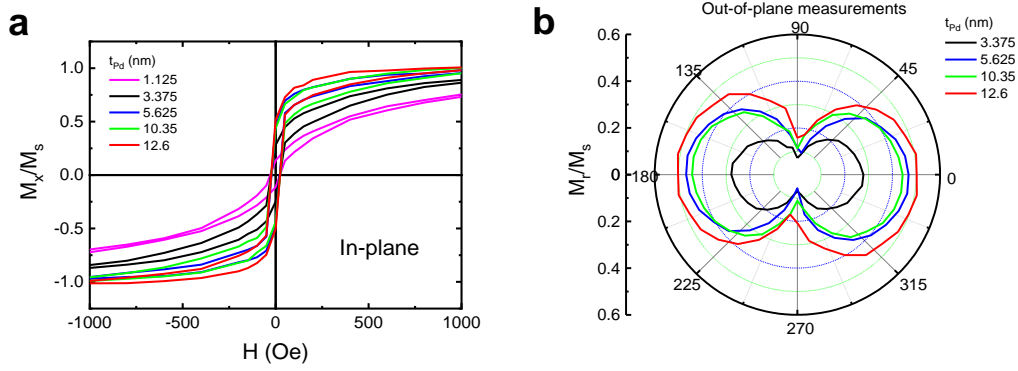

**Fig. S4.** (a) In-plane hysteresis loops for Ru-series. (b) Polar diagrams of $M_r/M_s=f(\varphi)$ for out-of-plane measurements.

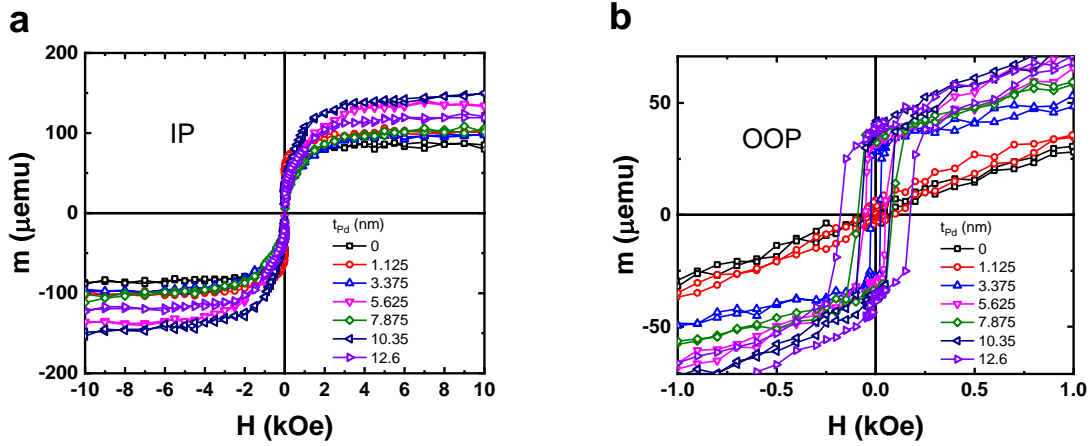

**Fig. S5**. Magnetic hysteresis loops for (a) in-plane (IP) and (b) out-of-plane (OOP) configurations for MgO-series with different $t_{Pd}$.

To define the effective saturation magnetization and MDL of our samples, we have fabricated series of samples with different thicknesses of CoFeSiB layer (from 0.6 to 2 nm) keeping all other parameters to be the same. The MDL values were defined from the linear slopes of $M_s t_{FM} = f(t_{FM})$ dependences, where $t_{FM}$ is the nominal thickness of the FM layer, as the intersection of the approximating lines with the $x$-axis. Initially, $M_s$ values were calculated as a ratio of the magnetic moment at saturation ($M$) to a unit volume ($V$) of the FM layer with $t_{FM}$. Then, the effective saturation magnetization was extracted as a slope of the $M_s t_{FM} = f(t_{FM})$ lines. The determined values are 927±20, 943±20 and 710±15 emu/cm$^3$ for MgO-, Ta- and Ru-series, respectively.

## 4. Approximation of the interface roughnesses by sinusoidal function

The modulated surface roughness in our systems can be roughly approximated by a sinusoidal function.[7] Below there are two examples for samples with $t_{Pd}$ = 3.38 and 12.6 nm, Fig. 3. This approximation is only needed for an analysis of the possible surface topography and increased area contribution to the enhancement of PMA and DMI.

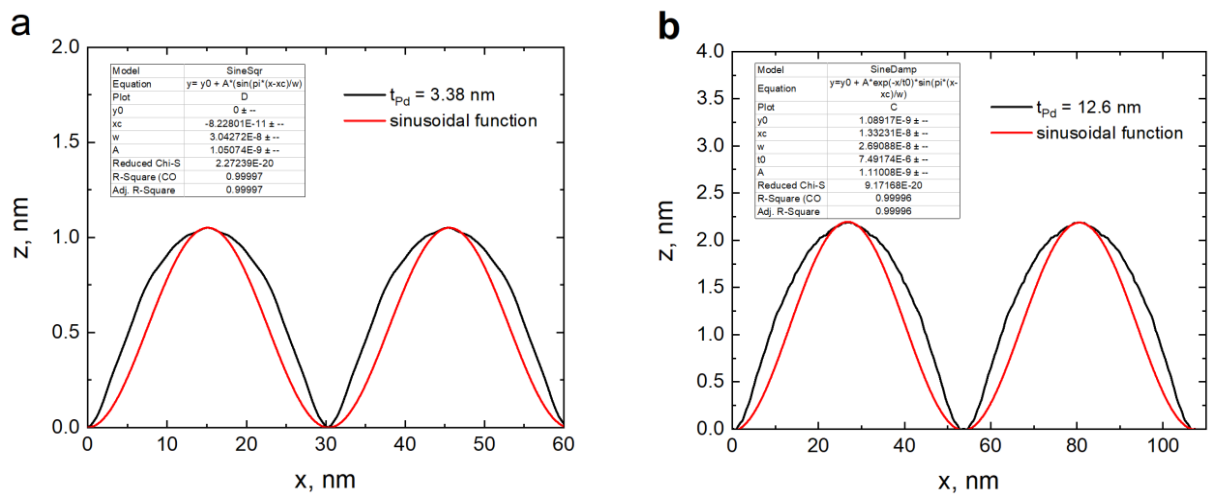

**Fig. S6.** Approximated experimental data (red curves) of the Pd surface roughness for samples with $t_{Pd}$ = 3.38 (a) and 12.6 nm (b). Experimental profiles (black curves) of the wavy roughness were taken from STM images of the Pd surface.

## 5. X-ray diffraction analysis

In our case we have epitaxially grown Pd(111) seed layers, which have epitaxial (single crystal) structure without formation of grains. The prepared Pd/Pt stacks have coherent interfaces (because of identical crystalline structure and small lattice mismatch) and can be considered as a superlattices[8]. The epitaxial Pd seed layer allows to exclude any possible effect of crystallinity on the consequently grown layer of Pt. The Pd layers have perfect planes of single crystal Pd(111) without crystallites formation. It confirms the same high surface quality of Pd(111) with its increasing thickness.

We have done XRD analysis of two series of samples used in the paper, Fig. S7. Since the thickness of Pt is just 2 nm and the reference peak positions of Pt(111) (39.8°) and Pd(111) (40.1°) are almost the same, it is practically impossible to indexing the Pt(111) peak. The main peak near 40˚ mainly corresponds to Pd, this is why the peak is observed more clearly as the thickness of Pd increases. We have had empirically found oscillation peaks such as seen for $t_{Pd}$=12.6 nm during measuring a single crystal film. These oscillations could be a result of a little different measurement conditions (mirror, beam or slit). Thus, we cannot conclude about any Pd-thickness-driven shift of Pt(111) peaks for both series of samples. But if take into account the influence of Ta in series 1 and Ru and Ta in series 2, we can claim that Pd(111) peaks are not shifted with increasing thickness proven the same crystalline quality of Pd(111) and, consequently, Pt(111) for all samples with various thicknesses.

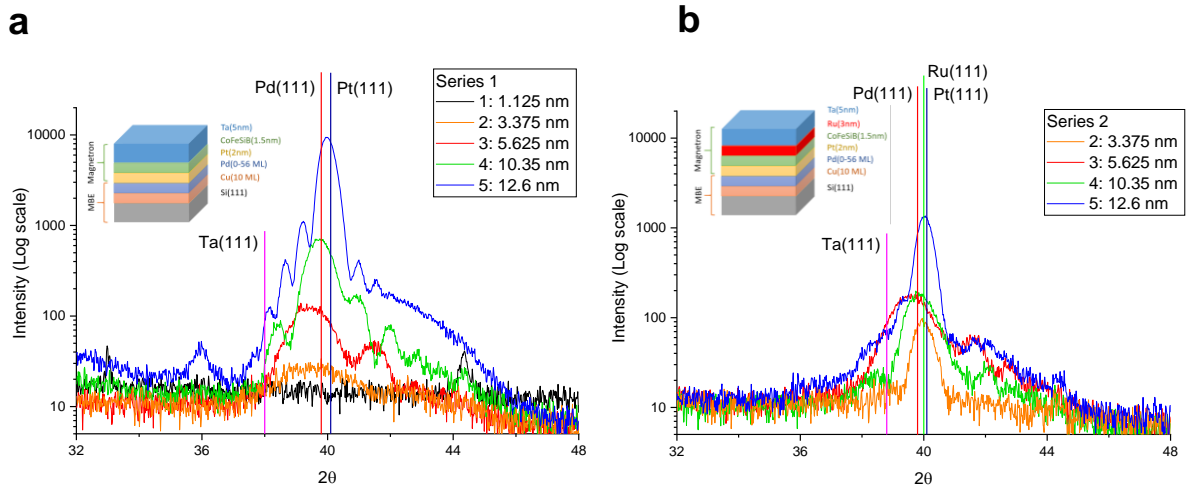

**Fig. S7.** XRD spectra analysis of samples from series 1 (a) and 2 (b) with different Pd thicknesses.